\documentclass[aps,pra,reprint,amssymb,amsmath, superscriptaddress,noeprint]{revtex4-2} 

\usepackage{graphicx}
\usepackage{dcolumn}
\usepackage{bm}
\usepackage[colorlinks=true,allcolors=metalblue]{hyperref}
\usepackage{orcidlink}
\usepackage{physics}
\usepackage[dvipsnames,table]{xcolor}
\usepackage[skip=4pt]{caption}
\usepackage{subcaption}
\usepackage{tikz}
\usepackage{float}

\definecolor{metalblue}{HTML}{397378}

\begin{document}

\title{Meson spectroscopy of exotic symmetries of Ising criticality in Rydberg atom arrays} 

\author{Joseph Vovrosh\,\orcidlink{0000-0002-1799-2830}}
\email{joseph.vovrosh@pasqal.com}
\let\comma,
\affiliation{PASQAL SAS, 24 rue Emile Baudot - 91120 Palaiseau,  Paris, France}

\author{Julius de Hond\,\orcidlink{0000-0003-2217-934X}}
\let\comma,
\affiliation{PASQAL SAS, 24 rue Emile Baudot - 91120 Palaiseau,  Paris, France}

\author{Sergi Julià-Farré\,\orcidlink{0000-0003-4034-5786}}
\let\comma,
\affiliation{PASQAL SAS, 24 rue Emile Baudot - 91120 Palaiseau,  Paris, France}

\author{Johannes Knolle\,\orcidlink{0000-0002-0956-2419}}
\let\comma,
\affiliation{Technical University of Munich, TUM School of Natural Sciences,
Physics Department, TQM, 85748 Garching, Germany}
\affiliation{Munich Center for Quantum Science and Technology (MCQST), Schellingstr. 4, 80799 M\"unchen, Germany}
\affiliation{Blackett Laboratory, Imperial College London, London SW7 2AZ, United Kingdom}

\author{Alexandre Dauphin\,\orcidlink{0000-0003-4996-2561}}
\email{alexandre.dauphin@pasqal.com}
\let\comma,
\affiliation{PASQAL SAS, 24 rue Emile Baudot - 91120 Palaiseau,  Paris, France}

\begin{abstract}
The Ising model serves as a canonical platform for exploring emergent symmetry in quantum critical systems. The critical point of the 1D Ising chain is described by a conformal Ising field theory, which remains integrable in the presence of a magnetic perturbation, leading to massive particles associated with the exceptional Lie algebra $E_8$. Weakly coupling two Ising chains into a ladder breaks this integrability and is predicted to confine the elementary excitations of each chain into a richer spectrum of bound states organized by a $\mathcal{D}^{(1)}_8$ symmetry. Experimental signatures of $E_8$ excitations have arguably been observed in scattering studies of the spin chain material CoNb$_2$O$_6$, but direct evidence of confinement driven by inter-chain coupling has remained elusive. Here, we probe these emergent symmetries in a Rydberg atom quantum processing unit, leveraging its tunable geometry to realize both chain and ladder configurations. We identify mass spectra consistent with $E_8$ at the single-chain critical point and, in the weakly coupled ladder, report the first signatures of confinement of Ising excitations into the bound-state spectrum predicted by $\mathcal{D}^{(1)}_8$ symmetry. Our results demonstrate the power of Rydberg platforms for investigating symmetry emergence in quantum many-body systems and provide a direct window into the interplay of confinement, geometry, and criticality.
\end{abstract}

\maketitle

\begin{figure*}
    \centering    
    \includegraphics[width=0.8\textwidth]{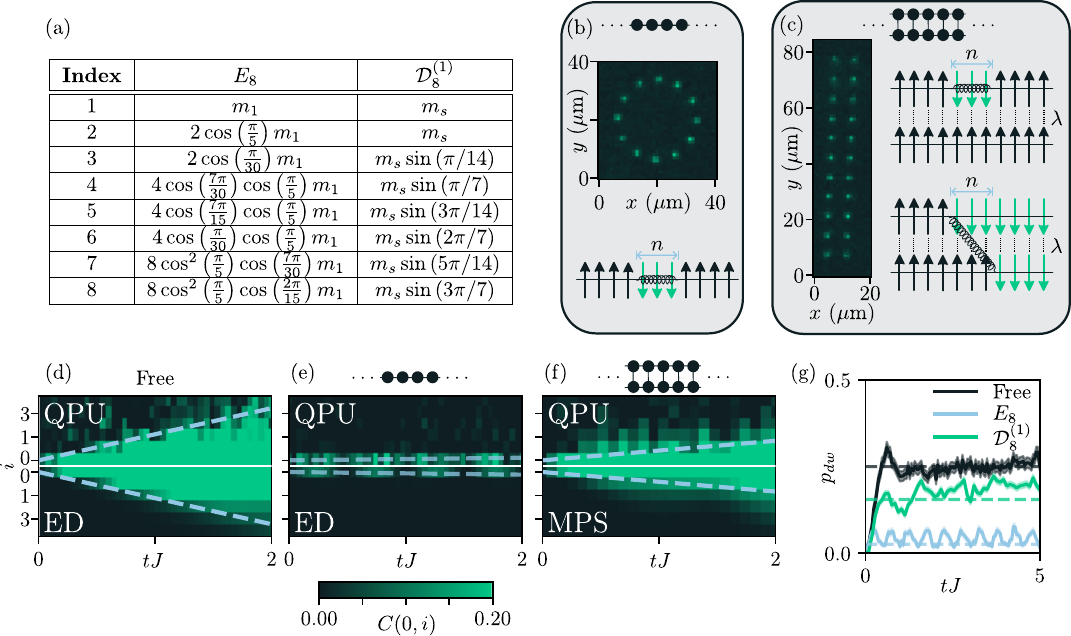}
    \caption{(a) Meson mass spectra predicted by emergent $E_8$ and $\mathcal{D}^{(1)}_8$ symmetries \cite{zamolodchikov1989integrals, LeClair_1998}. (b-c) Atomic register geometries implemented on the neutral-atom platform to realize (b) a 1D Ising chain and (c) a quasi-1D Ising ladder, together with schematics of the corresponding domain-wall meson excitations. (d-f) analogue QPU measurements (top) and exact numerical results (bottom) for correlation spreading following a quench from the $\ket{00\cdots00}$ state. Panels (d-e) show the Ising chain with $N=12$, $h_x=1$: (d) $h_z=0$ and (e) $h_z=4$. Panel (f) shows weakly coupled Ising ladders with $N=11$, $h_z=0$, and interchain coupling $\lambda=1$. Dashed blue lines indicate theoretical light-cone velocities extracted from meson dispersion relations. (g) Analogue QPU measurements of the domain-wall density $p_{dw} = (1-\langle \bar{ZZ}\rangle)/2$, where the average is taken over all nearest-neighbor pairs. Horizontal lines denote the long-time average domain-wall number obtained from classical simulations.}
    \label{Fig1}
\end{figure*}

\textit{Introduction} - Emergent symmetries at quantum critical points offer profound insights into the universal behavior of  many-body systems~\cite{mussardo2010statistical}. Among these, the exceptional Lie algebra $E_8$ has garnered significant attention because it emerges at the quantum critical point of the basic one-dimensional (1D) transverse field Ising chain (TFIC) when subjected to a longitudinal field~\cite{zamolodchikov1989integrals, fonseca2001isingfieldtheorymagnetic, fonseca2006isingspectroscopyimesons}. Instead of a quantum critical continuum response, discrete peaks of bound states appear in the structure factor, signatures of which have been experimentally observed in the quasi-1D Ising ferromagnet CoNb$_2$O$_6$~\cite{Coldea_2010}. Remarkably, the spectrum of bound states have universal energy ratios dictated by the $E_8$ symmetry given in Fig.~\ref{Fig1}(a).

However, subsequent studies have indicated that the excitation spectrum of CoNb$_2$O$_6$ may be more accurately described by the affine Lie algebra $\mathcal{D}_8^{(1)}$~\cite{LeClair_1998, gao2024spindynamicsdarkparticle,xi2024emergent}, arising from strong interchain interactions (see Fig.~\ref{Fig1}(a)) that effectively render the system as an Ising ladder rather than a simple chain. This shift in symmetry underscores the sensitivity of emergent phenomena to the underlying geometry and interactions within the system.

The bound states emerging at perturbed quantum critical points are intimately related to the phenomenon of confinement, i.e.\ deep in the ordered phase perturbations lead to a monotonically increasing potential between domain-wall quasi-particles and the formation of discrete mesonic bound states~\cite{Rutkevich_2008,greensite2011introduction}. The latter then continuously connect to the discrete particles of the critical points and govern the real-time post-quench dynamics~\cite{Kj_ll_2011,Kormos_2016}. In quantum Ising chains confinement can be induced through various mechanisms, including external magnetic fields~\cite{PhysRevD.18.1259, Kormos_2016, Vovrosh_2021}, long-range interactions~\cite{Liu_2019, Tan_2021} and interchain couplings~\cite{PhysRevB.102.014426, Lagnese_2020}, leading to rich excitation spectra that reflect the system's symmetry properties~\cite{Kj_ll_2011,Lamb_2024}.

In this work, we utilize the versatility of Rydberg atom arrays to experimentally investigate the emergence of $E_8$ as well as, for the first time, $\mathcal{D}_8^{(1)}$ symmetries in 1D Ising chains and Ising ladders, respectively~\cite{Browaeys_2020,Schauss_2018,Labuhn_2016}. By precisely engineering the geometry and interactions within these quantum simulators, we are able to probe the critical points of these systems and observe the corresponding bound state spectra. While signatures of confinement have previously been observed in isolated Ising chains, both in analogue~\cite{Tan_2021} and digital quantum processing units (QPUs)~\cite{Vovrosh_2021,Vovrosh_2022}, such phenomena have not yet been explored in Ising ladder geometries within quantum simulators. reveal clear signatures of confinement, together with characteristic frequencies that align with the predictions of the $E_8$ and $\mathcal{D}_8^{(1)}$ theories. This provides the first experimental evidence of confinement physics in an Ising ladder implemented on a quantum simulator, and establishes the emergence of these symmetries at the corresponding critical points. Our study demonstrates the power of QPUs in the analog mode in exploring complex many-body physics and paves the way for future investigations into higher-dimensional systems and other exotic symmetries.

\textit{Confinement in Ising Chains} - Throughout this work, we consider the transverse field Ising model, which describes interacting spin-$\frac{1}{2}$ particles\begin{equation}\label{eq:ising_hamiltonian}
    H_I = -J\sum_{i,j}\sigma_{i}^z\sigma_{j}^z-h_x\sum_i\sigma_{i}^x,
\end{equation}
where $\sigma_{i}^\alpha$ are the usual Pauli matrix operators acting on the $i^{\text{th}}$ site, $J$ is the spin-spin interaction, and $h_x$ is the transverse field. In the rest of the work, we set $J=1$.

In a one-dimensional TFIC quasiparticle excitations of this Hamiltonian correspond to free fermions. By considering $h_x < h_c$, where $h_c = 1$ is the critical point of the Hamiltonian, quasiparticles are well approximated by simple domain walls. Introducing a finite longitudinal field, $-h_z\sum_i\sigma_{i}^z$, two quasiparticles experience a linear confining force proportional to both their separation and $h_z$~\cite{Rutkevich_2008}, see Fig.~\ref{Fig1}(b). Consequently, each pair of quasiparticles form a mesonic bound state with a mass defined by $m_i = E_i-E_0$ where $E_i$ is the $i^{\text{th}}$ energy level (with zero momentum). Confinement is known to slow correlation and entanglement spread \cite{Vovrosh_2021}, delay thermalization \cite{Liu_2019, Birnkammer_2022} and lead to interesting phenomena such as string breaking~\cite{PhysRevB.102.014308} and inelastic collisions~\cite{Vovrosh_2022, karpov2020spatiotemporaldynamicsparticlecollisions,milsted2012collisions,jha2024real, schuhmacher2025observationhadronscatteringlattice,Surace_2021,Bennewitz_2025}. 

Confinement is also known to occur in the weakly coupled Transverse Field Ising Ladder (TFIL) for $h_x < h_c$, and \textit{without} the need for an additional magnetic field, 
\begin{equation}
    H = H_I^1+H_I^2 - \lambda\sum_i\sigma_{1,i}^z\sigma_{2,i}^z,
\end{equation}
in which $H^i_I$ acts on the $i^{\text{th}}$ chain, $i\in\{1,2\}$, and $\lambda\ll 1$. In this regime, pairs of excitations also experience a linear confining force, this time proportional to their separation and to the interchain coupling, $\lambda$. Recent theoretical studies employing time-dependent density matrix renormalization group methods have revealed that the excitation spectrum of the TFIL exhibits a rich structure of bound states~\cite{PhysRevB.102.014426}. These include both {\it intra} and {\it interchain} mesonic excitations, as well as bound states of mesons themselves, see Fig.~\ref{Fig1}(c). The presence of these higher-order bound states indicates a hierarchical confinement mechanism unique to the ladder geometry described by two coupled Ising conformal field theories~\cite{LeClair_1998}.

Although the quantum field theory predictions associated with both the $E_8$ and $\mathcal{D}^{(1)}_8$ symmetries for these models are derived assuming only nearest-neighbor interactions for both the TFIC with a longitudinal field and TFIL, the $E_8$ symmetry remains approximately preserved even in the presence of long-range interaction tails and other erroneous terms, such as those encountered in trapped-ion systems~\cite{PhysRevA.105.022616} and additional subdominant terms required for more precise models of CoNb$_2$O$_6$~\cite{Kj_ll_2011}. In the following sections, we extend this perspective to show that the $\mathcal{D}^{(1)}_8$ symmetry is robust against long-range interactions, such as those realized in Rydberg arrays.

\textit{Experimental Considerations} - An Ising-type Hamiltonian can be naturally realized with arrays of neutral atoms in optical tweezers. When excited to high-lying Rydberg states, they  exhibit large dipolar interactions. By defining the atomic ground state as $\ket{0}$ and the Rydberg state as $\ket{1}$, the Hamiltonian can be expressed as 
\begin{equation}\label{eq:rydberg_hamiltonian}
H_{\text{Ryd}}=\sum_{i,j > i}\frac{C_6}{|\mathbf{r}_i-\mathbf{r}_j|^6}n_in_j+\frac{\hbar\Omega(t)}{2}\sum_i \sigma^x_i-\sum_i\frac{\hbar\delta_i(t)}{2} \sigma^z_i,
\end{equation}
where $n_i=(1+\sigma^z_i)/2$, $C_6>0$ is the strength of the Van der Waals antiferromagnetic interaction, and $\Omega(t)$ and $\delta(t)$ are the Rabi frequency and detuning induced by the external laser~\cite{Henriet_2020}. By adjusting both the Rabi frequency and detuning, one can map this Hamiltonian to an Ising model exactly for systems with periodic boundary conditions (PBC), and up to a local longitudinal field at the edge of the system for open boundary conditions (OBC), full details can be found in App.~\ref{appendix_ryd}. Furthermore, we can also probe the dynamics of the ferromagnetic Ising model for initial states preserving time-reversal symmetry~\cite{menu2020gaussian}.

In particular, a TFIC with a longitudinal field can thus be realized in a Rydberg atom platform through arranging atoms in a 1D chain, or in our case, a circular geometry allowing for the consideration of PBC, see Fig.~\ref{Fig1}(b). Furthermore, we realize the TFIL in our Rydberg atom platform through arranging atoms in two 1D chains. Note, given a two-dimensional (2D) trapping region, it is not possible to exactly realize an Ising ladder with PBC and thus throughout this work we consider OBC for this quasi-1D system, see Fig.~\ref{Fig1}(c).

\textit{Observations of Confinement} - A quintessential signature of confinement dynamics is the suppression of correlation spreading following a sudden quench, that is $C(i,j) = \langle\sigma_i^z\sigma_j^z\rangle - \langle\sigma_i^z\rangle\langle\sigma_j^z\rangle$. In conventional critical systems, correlations typically spread ballistically, governed by the maximal group velocity of quasiparticles, for the TFIC this is given by $\mathcal{V}=2\min{(1,h_x)}$~\cite{calabrese_time_2006, Kormos_2016, lieb_finite_1972}. However, in confined systems, the presence of an effective potential between quasiparticles significantly slows down their propagation, leading to a characteristic suppression of entanglement growth and a deviation from the expected light-cone dynamics. This behavior has been extensively studied in both theoretical models~\cite{Kormos_2016, Liu_2019, PhysRevB.102.014426} and experimental platforms~\cite{Tan_2021, Vovrosh_2021, kebrič2024confinement11dmathbbz2lattice}, making it a key observable for identifying confinement effects. 

In our experiments, we measure the connected correlation function $C(i,j)$ following a quantum quench from the $\ket{00\dots0}$ state, allowing us to directly track the propagation of correlations across the system. The experimentally obtained correlation data show strong quantitative agreement with classical numerical simulations across all explored regimes, up to the experimental noise, see Fig.~\ref{Fig1}(d-f). Furthermore, we directly observe the effect of confinement by comparing different regimes. In the absence of confinement, correlations spread with a clear light-cone, consistent with the Lieb-Robinson bound analytically known for the TFIC with $h_z=0$, $v=2J\min(1,h_x)$~\cite{Kormos_2016, lieb_finite_1972}, which governs the maximum speed of information propagation in quantum systems, see Fig.~\ref{Fig1}(d). However, in the presence of a longitudinal field, or for weakly coupled Ising chains, the spreading of correlations is significantly slowed, see Fig.~\ref{Fig1}(e-f). 

This behavior aligns with theoretical predictions for confined systems, where the effective meson velocity, calculated through the maximal group velocity of the low lying energy level (see App.~\ref{appendix_velocities}), governs the dynamics of correlation spreading~\cite{Vovrosh_2021}. However, this prediction is not exact: in the TFIL regime, correlations are observed beyond the light cone defined by the meson velocity in both the analogue QPU and classical data, indicating that the true Lieb–Robinson velocity of the system is larger. The meson velocity therefore captures the primary confined propagation front, but does not fully determine the ultimate bound on information spreading. Beyond this, small additional fluctuations outside the light cone can also arise from experimental noise.

Beyond slowing the spread of correlations, we also observe that confinement suppresses the growth in the number of domain walls, a hallmark of constrained dynamics, see Fig.~\ref{Fig1}(g) right panels. This suppression is consistent with the picture of meson-like bound states limiting the available phase space for domain wall proliferation. These results provide direct experimental evidence of confinement-induced slowdown in both correlation spreading and excitation dynamics—a qualitative feature also observed in quantum simulations of confined lattice gauge theories in one~\cite{mildenberger2025confinement} and two dimensions~\cite{cochran2024visualizing}.

\begin{figure}
    \centering
    \includegraphics[width=0.4\textwidth]{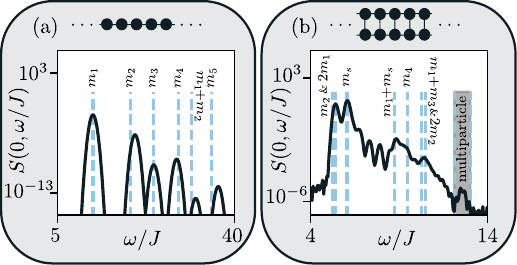}
    \caption{Zero-momentum dynamical structure factor for (a) the TFIC with a longitudinal field with $h_x = 1$ and $h_z = 4$, and (b) the TFIL with $h_x = 1$, $h_z = 0$, and inter-chain coupling $\lambda = 1$. In both cases, vertical dashed lines indicate the theoretically predicted mass ratios associated with emergent symmetries: $E_8$ in the chain and $\mathcal{D}^{(1)}_8$ in the ladder.}
    \label{Fig2}
\end{figure}

\begin{figure*}
    \centering    
    \includegraphics[width=0.8\textwidth]{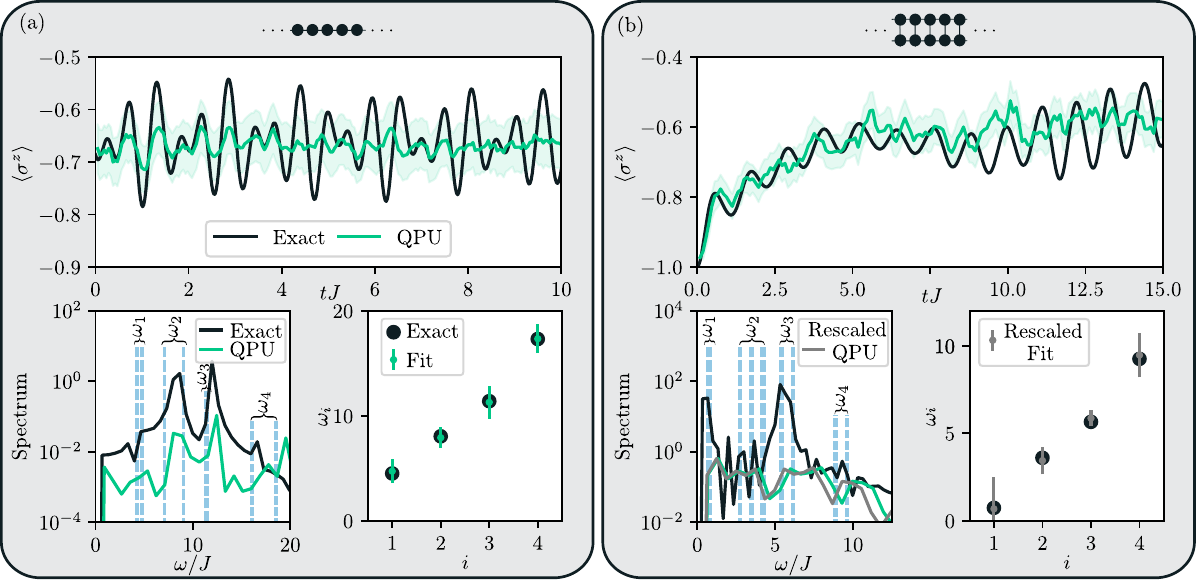}
    \caption{Analysis of post-quench dynamics performed on from the $\ket{00\cdots0}$ state to both the TFIC with a longitudinal field with (a) $N=12$, $h_x=1$ and $h_z=4$ and (b) the TFIL with $N=11$, $h_x=1$, $h_z=0$ and $\lambda=1$. The upper panels shows real-time data for $\sigma^z$ for both analogue QPU, in which a moving average has been used to reduce statistical fluctuations, and classical simulations. The bottom left panel shows the Fourier transformation of this data, with the dominant peaks, corresponding to the linear combinations of the meson masses predicted by $E_8$ and $\mathcal{D}^{(1)}_8$ symmetry - and their differences - are indicated by vertical dashed lines. The lower right panel shows a comparison between the dominant frequencies of the Fourier analysis of classical data and the fitted frequencies of the ansatz fitted to the QPU data. In (b) lower right panel, the frequencies are obtained from a reescaled QPU spectrum, motivated by residual experimental offsets (see App. D).}
    \label{Fig3}
\end{figure*}

\textit{Observation of Symmetries} - The hallmark of the emergent $E_8$ symmetry is the appearance of eight massive excitations whose mass ratios are universal and fixed by the integrable field theory describing the Ising chain at critical transverse field perturbed by a longitudinal field. The exact solution predicts a complete mass spectrum uniquely determined by the $E_8$ Lie algebra structure, with ratios that are independent of microscopic details. These universal ratios therefore provide a direct “smoking-gun” signature of the symmetry. In practice, we extract the excitation gaps from the peak positions of the zero-momentum dynamical structure factor (DSF) $S(k=0,\omega/J)$, whose peak frequencies $\omega_i$ correspond to the meson masses $m_i$. The ratios $\omega_i/\omega_1$ can then be directly compared to the theoretical predictions~\cite{Coldea_2010,Kj_ll_2011}.

A closely analogous situation arises for the ladder geometry, whose scaling limit is described by the integrable field theory associated with the $\mathcal{D}^{(1)}_8$ algebra. This theory likewise predicts a universal set of massive excitations with fixed mass ratios. However, as discussed in Refs.~\cite{gao2024spindynamicsdarkparticle,xi2024emergent}, symmetry constraints imply that not all predicted modes contribute with visible weight to the zero-momentum dynamical structure factor. As a result, only a subset of the theoretical masses appear as peaks in $S(k{=}0,\omega)$. Taking these selection rules into account, the observed peak structure is consistent with the universal $\mathcal{D}^{(1)}_8$ predictions.

To study the emergent $E_8$ and $\mathcal{D}^{(1)}_8$ symmetries at the critical points of the Ising chain and ladder, we first optimize the atomic register, Rabi frequency, and detuning to ensure that the systems under investigation accurately reflect the theoretically predicted energy spectra. In order to verify the presence of an $E_8$ symmetry, we examine a 12-qubit Ising chain with PBC. By tuning the Rabi frequency and detuning such that $h_x = 1$ and $h_z=4$ we obtain the resulting $S(k=0,\omega/J)$ from the full eigenspectrum obtained via exact diagonalization, as shown in Fig.~\ref{Fig2}(a). The ratios of the peaks seen in these results closely match the theoretical predictions for $E_8$ symmetry. Note, that despite the relatively small length of the chain the lowest masses follow the prediction because the strong longitudinal field suppresses finite size effects~\cite{PhysRevA.105.022616}.

Similarly, to assess whether the Rydberg ladder is well described by the $\mathcal{D}^{(1)}_8$ symmetry, we consider two 11-qubit Ising chains with OBC. The atomic separation, Rabi frequency, and detuning are adjusted such that $\lambda = 1$, $h_x = 1$, and $h_z = 0$, with a total of 22 qubits. We simulate the DSF with the help of matrix product state methods~\cite{bidzhiev2025efficientemulationneutralatom}. As shown in Fig.~\ref{Fig2}(b), the peaks in $S(k=0,\omega/J)$ align well with the predicted peaks for the $\mathcal{D}^{(1)}_8$ symmetry.

To observe both emergent symmetries, $E_8$ and $\mathcal{D}^{(1)}_8$, on the analogue QPU we extract meson spectra through real-time data: we first perform a quench in each system and measure the resulting fluctuations in the local magnetization, $\langle\sigma_i^z\rangle$. By subsequently performing a Fourier transform on the post-quench dynamics of $\langle\sigma_i^z\rangle$, we obtain a clear spectral decomposition with dominant frequencies, ${\omega_i}$, that can be matched to linear combinations of the theoretically predicted meson mass ratios. Secondly, we can refine the determination of the peak positions by fitting a Fourier ansatz directly to the analogue QPU data. This procedure allows for a more precise extraction of the dominant frequencies, which we then compare to the theoretical meson mass ratios.

For the TFIC with a longitudinal field, the meson spectrum spans a wide energy range, so we select a relatively large longitudinal field, $h_z = 4$, which allows us to compress multiple relevant frequencies into the timescales obtainable by the quantum hardware's operational regime, $tJ=O(10)$, significantly exceeding those achieved in previous works \cite{Tan_2021, Vovrosh_2021_mit, Lamb_2024}. However, this large field causes the trivial initial state $\ket{0}^{\otimes N}$ to primarily couple to the lowest masses, $m_1$ and $m_2$, of the $E_8$ spectrum, severely limiting the visibility of higher meson states. To address this, we employ a two-stage quench protocol: the system is first evolved under a ramp protocol from $h_x = 1$ and $h_z = 0$ to the desired and $h_z = 4$ over a $400$ns period of time. This generates an intermediate state with nontrivial structure with an enhanced overlap with multiple $E_8$ meson eigenstates. We then evolve this state with $h_x = 1$ and $h_z = 4$. The real-time data collected by the analogue QPU show remarkable agreement to the exact results obtained with classical methods, with deviations attributable to experimental noise in the analogue QPU data. These can then be used to obtain a rich frequency spectrum, see Fig.~\ref{Fig3}(a).

In contrast, for the $\mathcal{D}^{(1)}_8$ case realized in the TFIL geometry, the ground state $\ket{0}^{\otimes 2N}$ naturally overlaps with several meson modes. As such, a single quench from this initial state is sufficient to excite a broad spectrum of meson masses, eliminating the need for a preparatory evolution stage. Hence we consider a single quench from $\ket{0}^{\otimes N}$ to the TFIL with $h_x=1$ and $\lambda=1$, see Fig.~\ref{Fig3}(b). The dynamics of the TFIL performed here are highly sensitive to the precise choice of Hamiltonian parameters, such that even small miscalibrations of the analogue QPU lead to noticeable shifts in the observed oscillation frequencies. Nevertheless, because these frequencies are constrained by the emergent $\mathcal{D}^{(1)}_8$ symmetry, their ratios remain robust against global calibration offsets. Crucially, the high level of control and quantitative understanding of noise in our device allows us to identify the dominant sources of miscalibration and their impact on the overall energy scale. Since such imperfections primarily induce a uniform renormalization of the spectrum, without altering the underlying symmetry, imposed structure—we can leverage them as a diagnostic tool to independently extract the effective energy shift. This, in turn, enables a controlled rescaling of the experimental time axis, bringing the analogue QPU data into quantitative agreement with exact theoretical predictions while preserving the underlying physics (see App.~\ref{appendix_noise} for details).

In order to refine the observed frequencies in the analogue QPU data, we fit the resulting time dynamics from both the TFIC and TFIL simulations to a multi-frequency Fourier ansatz. We choose the same number of dominant frequenices, $\omega_i$ as predicted by the underlying symmetries and fit using the ADAM optimizer~\cite{kingma2017adammethodstochasticoptimization}. We run 20 independent fits over a maximum of 20,000 epochs each to ensure convergence and robustness (see App.~\ref{appendix_foruier}). The extracted frequencies are then averaged across fits. The results of both protocols are shown in Fig.~\ref{Fig3} (lower panels). Despite the presence of statistical fluctuations and hardware-induced noise, the frequencies of the analogue QPU data agree well with the theoretical predictions $\omega_i$. The agreement to classical simulations thus confirms that the analogue QPU captures the key spectral signatures associated with both $E_8$ and $\mathcal{D}^{(1)}_8$ symmetry. We stress that the confinement mechanism for the two cases is quite distinct because the TFIL does not need a longitudinal field, but rather the other chain acts as a dynamical confining field captured in our ladder geometry. For a study of the noise seen in the analogue QPU results, see App.~\ref{appendix_noise}.

\textit{Conclusion and Outlook} - In this paper, we observed emergent symmetries of Ising models at criticality with the help of a neutral atom QPU in analogue mode. We first probed confinement for both a 1D Ising chain with PBC and an Ising ladder with OBC by taking advantage of the flexibility of the atomic register. In particular, we observed the slow down of correlation spread due to confinement in these models. Finally, by performing meson spectroscopy, we showed that the emergent $E_8$ and $\mathcal{D}^{(1)}_8$ symmetries in these models can be extracted from the real time quench dynamics.  Our analogue QPU therefore allows us to probe the low-energy spectrum of a toy model capturing the essential magnetic physics of CoNb$_2$O$_6$, and we observe qualitative agreement with neutron scattering measurements.

While all the dynamics simulated on the analogue QPU in this work are classically tractable for the system sizes and dimensions considered, neutral atom QPUs in analogue mode can be scaled to hundreds of atoms in 2D. In fact, some studies have already studied the physics of confinement in 2D systems with neutral atom platforms~\cite{Gonz_lez_Cuadra_2025}, as well as superconducting qubits~\cite{cochran2024visualizing}. With improved initial state preparation, meson scattering experiments~\cite{Vovrosh_2022, karpov2020spatiotemporaldynamicsparticlecollisions,milsted2012collisions,jha2024real} will be possible  and will allow for a better characterization of the elastic or inelastic nature of the scattering.  

Finally, neutral atom QPUs in analogue mode can readily be used to probe Hilbert Space Fragmentation~\cite{Yoshinaga_2022}. The latter has been observed in 1D neutral atom platforms~\cite{PhysRevX.15.011035,karch2025probing} and going to 2D promises a much richer phenomenology. It is known that accurately simulating the dynamics of 2D models near criticality is exponentially costly in memory and computation time~\cite{osborne_hamiltonian_2012, coleman_introduction_2015} and therefore the analogue QPUs can shed a new light allow to quantum criticality in 2D. 

\acknowledgments
A.D. thanks Christophe Jurczak for a careful reading of the manuscript. J.K.\ thanks the hospitality of the Aspen Center for Physics, which is supported by National Science Foundation grant PHY-2210452. J.K.\ acknowledges support from the Deutsche Forschungsgemeinschaft (DFG, German Research Foundation) under Germany’s Excellence Strategy– EXC–2111–390814868 and DFG Grants No. KN1254/1-2, KN1254/2-1 and TRR 360 - 492547816, as well as the Munich Quantum Valley, which is supported by the Bavarian state government with funds from the Hightech Agenda Bayern Plus. J.K.\ further acknowledges support from the Imperial-TUM flagship partnership. Pasqal's team acknowledges funding from the European Union the projects EQUALITY (Grant Agreement 101080142) and PASQuanS2.1 (HORIZON-CL4-2022-QUANTUM02-SGA, Grant Agreement 101113690).

\bibliography{Biblio}

\clearpage
\begin{widetext}
\appendix

\section{Simulating the Ising model with Rydberg atoms}
\label{appendix_ryd}

Rydberg atom arrays provide a versatile and highly controllable platform for simulating quantum spin models, including variants of the transverse-field Ising model. In these systems, we load registers by using roughly $2N$ tweezer traps, which capture approximately $N$ atoms on average. These are imaged, after which an optimal series of moves is calculated, and then performed, to assemble the desired register. Prior to running the Rydberg excitation sequence, an image is taken to determine whether the rearrangement is successful. These atoms are then coherently excited to high-lying Rydberg states, where they experience strong van der Waals interactions. As such, each atom effectively behaves as a spin-1/2 system, with the computational basis defined by the ground state $\ket{g}$ and a Rydberg state $\ket{r}$. 

In this work, we employ \textsuperscript{87}Rb atoms, encoding the states as $\ket{g} = \ket{5S_{1/2}, F=2, m_F=2}$ and $\ket{r} = \ket{75S_{1/2}, m_J = 1/2}$, utilizing a two-photon excitation scheme via the $6P_{3/2}$ intermediate state, detuned by $400$MHz. The resulting dynamics of the system are governed by the many-body Hamiltonian~\cite{Browaeys_2020, Henriet_2020}:
\begin{equation}
H_{\text{Ryd}} = \sum_{i,j > i} \frac{C_6}{|\mathbf{r}_i-\mathbf{r}_j|^6} n_i n_j + \frac{\Omega}{2} \sum_i \sigma_i^x - \frac{\delta}{2} \sum_i \sigma_i^z,
\end{equation}
where $n_i = (1 + \sigma_i^z)/2$ is the projector onto the Rydberg state, $r_{ij} = |\mathbf{r}_i-\mathbf{r}_j|$ is the interatomic distance, $C_6/2\pi = 1948~\mathrm{GHz\,\mu m^6}$ is the van der Waals interaction coefficient, $\Omega$ is the Rabi frequency of the laser driving the $\ket{g} \leftrightarrow \ket{r}$ transition, and $\delta$ is the detuning of the laser. By making the mapping 
\begin{equation}
J_{ij} = \frac{C_6}{4r_{ij}^6},\;\;\;\;\;
h_x = \frac{\Omega}{2},\;\;\;\;\;
h_z = \frac{\delta}{2},
\end{equation}
we obtain an effective spin-1/2 Hamiltonian resembling the transverse-field Ising model
\begin{equation}
H_{\text{Ryd}} = \sum_{i,j > i} J_{ij} \sigma_i^z \sigma_j^z + h_x \sum_i \sigma_i^x + h_z \sum_i \sigma_i^z + \sum_i\Delta_{i},
\label{eq:ising_rydberg}
\end{equation}
such that
\begin{equation}
\Delta_{i} = \sum_{j>i} J_{ij}(\sigma_i^z + \sigma_j^z).
\end{equation}
This erroneous term acts as a site dependent, effective shift in the longitudinal field. In systems with periodic boundary conditions, this contribution can be fully absorbed into a redefinition of the global detuning, i.e. $\delta_i = -2h_z+\sum_j J_{ij}/2$, thereby eliminating the effective longitudinal shift. While this cancellation is not exact for open boundary conditions, the long-range nature of the interaction, $J_{ij} \propto 1/r_{ij}^6$, ensures that the residual term is strongly localized near the system boundaries and rapidly becomes negligible in the bulk. Consequently, for the system sizes and geometries considered in this work, this boundary-localized correction does not affect the bulk physics or the dynamical regimes of interest.

Although $H_{\text{Ryd}}$ features only antiferromagnetic interactions ($J_{ij} > 0$), the dynamics of the ferromagnetic regime can still be explored. Specifically, the dynamics under $H_{\text{Ryd}}$ become equivalent (up to a global phase) to those under $-H_{\text{Ryd}}$, provided the initial state respects time-reversal symmetry. Throughout this work, we choose to set $\Omega/2\pi = 0.75~$MHz and $1.5~$MHz for simulations of the TFIC and TFIL respectfully; both the atomic separation and the detuning are then tuned to achieve the desired values of $h_x/J$ and $h_z/J$ in the bulk. In experiments, the Rydberg atom platform allows for site-resolved measurement of spin observables. In particular, the local magnetization $\langle \sigma_i^z \rangle$ and two-point correlations $\langle \sigma_i^z \sigma_j^z \rangle$ can be directly inferred from fluorescence imaging, which provides projective measurements of the atomic populations $n_i$ with single-site resolution.

\section{Meson velocities from DSF results}
\label{appendix_velocities}

\begin{figure}
    \centering
    \includegraphics[width=0.8\textwidth]{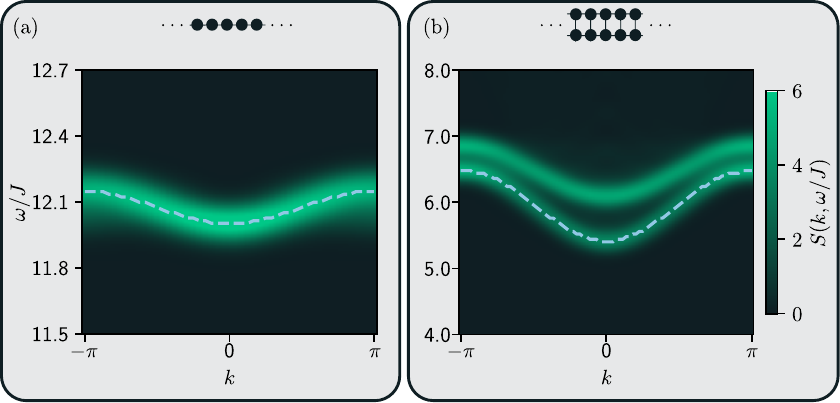}
    \caption{The dynamical structure factor for the two \textit{ferromagnetic} systems: (a) the TFIC with a longitudinal field with $h_x = 1$ and $h_z = 4$, and (b) the TFIL with $h_x = 1$, $h_z = 0$, and inter-chain coupling $\lambda = 1$. In both cases, dashed lines indicate the lowest energy level of the system.}
    \label{Sup_Fig1}
\end{figure}

In Fig.~\ref{Sup_Fig1}, we present the dynamical structure factor for both the \textit{ferromagnetic} transverse-field Ising chain and the transverse-field Ising ladder configurations considered in this work. The dynamical structure factor, which captures the spectral weight of excitations as a function of momentum and energy, reveals key features of the low-energy quasiparticle spectrum. From these spectra, we identify the dispersion relation of the lowest-lying excitations and extract the corresponding energy-momentum relationship, $\epsilon(k)$. The group velocity, defined as $v_g = \partial \epsilon(k) / \partial k$, characterizes the speed at which these quasiparticles propagate through the system. By locating the point at which this derivative is maximized, we determine the maximal group velocity, a key parameter that sets the upper bound for the speed of information and correlation spreading in the confined regime. 

From our analysis, we find that the maximal group velocity (in of units sites$\times J$) is approximately $v_g \approx 0.1$ for the TFIC with a longitudinal field and significantly larger, $v_g \approx 0.7$, for the TFIL. This difference reflects the varying strength of confinement in the two models, with the more weakly confined TFIL supporting faster, more delocalized quasiparticles. These values are consistent with the observed differences in the spreading of correlations and domain wall dynamics in the main text.

\section{Fourier Fitting Procedure}
\label{appendix_foruier}

To extract dominant frequency components from the analogue QPU data, we employ a Fourier fitting procedure based on the dominant spectral features observed in classical simulations. The fitting model is defined as
\begin{equation}
f(x) =  (c_0 e^{-\gamma x} + c_1) \left( a_0 + \sum_{i=1}^{N} a_i \cos(\omega_i x) + b_i \sin(\omega_i x) \right),
\end{equation}
where $c_i$, $\gamma$, $a_i$, and $b_i$ are real-valued coefficients to be optimized, $\omega_i$ are the frequencies to be fitted, and $\gamma$ is an exponential decay rate accounting for signal damping over time. We choose this Fourier form to isolate the dominant frequencies present in the analogue QPU data and enable direct comparison with the classical results. The fitting minimizes the mean squared error (MSE) between the analogue QPU data and the model, with all parameters --- including amplitudes, frequencies, decay rate, and offset --- allowed to vary unconstrained during optimization.

For each analogue QPU dataset, we initialize the frequencies $\omega_i$ to those obtained classically, as this initialization consistently leads to the lowest loss without overfitting. We perform 20 independent optimization runs using the ADAM optimizer implemented in PyTorch~\cite{kingma2017adammethodstochasticoptimization, paszke2019pytorchimperativestylehighperformance}, each over 20{,}000 epochs. The number of epochs was chosen because the loss typically plateaus by this point, indicating convergence. Finally, the frequencies extracted from each run are averaged to yield robust estimates that mitigate the effects of statistical fluctuations, hardware noise, and optimizer variability.

\section{QPU Hardware and Noise Analysis}
\label{appendix_noise}

\begin{figure}
    \centering
    \includegraphics[width=0.8\textwidth]{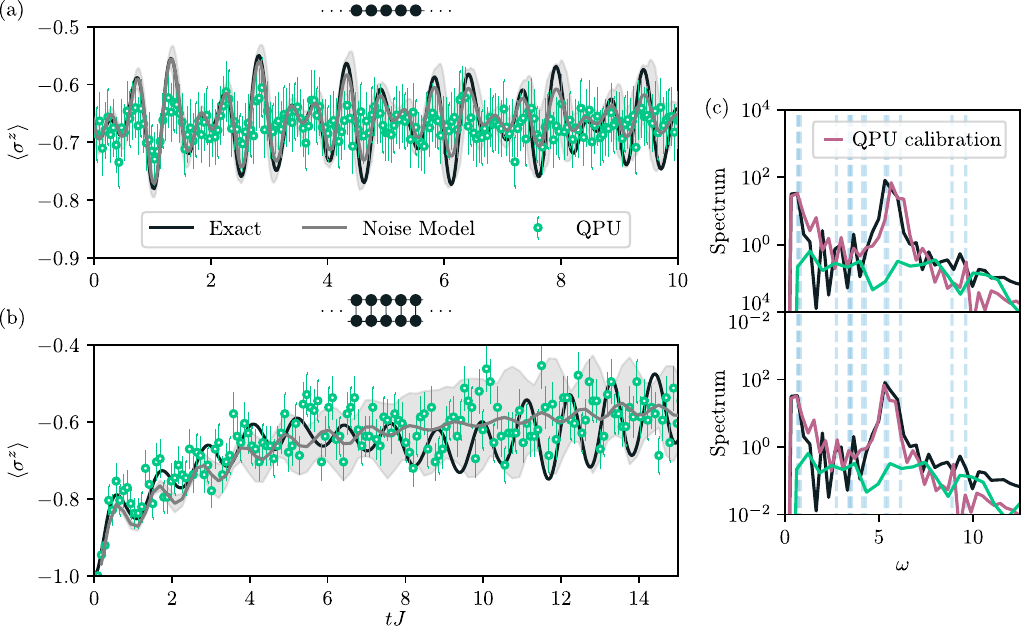}
    \caption{Noise analysis of the analogue QPU result presented in the main text. Here we show that by taking into consideration a experimentally realistic noise model we can recapture the results of the hardware. (a-b) Comparison of the noise model with the raw analogue QPU results for the TFIC and TFIL respectively. The noise model curve corresponds to the mean over 50 Monte Carlo trajectories, and the uncertainty shaded area corresponds to a $70\%$ confidence interval.  (c) The effect of rescaling the time axis of the analogue QPU data for the TFIL simulations. Note, we also include the classical simulation of the miscalibrated analogue QPU.}
    \label{Sup_Fig2}
\end{figure}

In this section, we analyze the experimental results presented in Fig.~\ref{Fig3} of the main text and compare them to the exact theoretical predictions. The experimental runs were carried out on our R$\&$D device FM1 operating at a shot rate of $\sim1$ Hz. For each simulation, observables were measured at 200 distinct time steps, with 250 shots taken per time step. Additionally, each batch of 250 shots required the realization of a custom atomic register, which introduced an overhead. Taking all these factors into account, the total runtime for each complete simulation was approximately 1 day. Multiple simulations were conducted with all yielding consistent performance across runs, which demonstrates reproducibility across independent runs.

In our numerical modeling of the analogue QPU, we incorporate the dominant experimental noise sources using independently calibrated parameters. Laser fluctuations, thermal motion of the atoms, and coupling to the environment give rise to Hamiltonian fluctuations, effective decay and dephasing channels, as well as state-preparation-and-measurement (SPAM) errors, as characterized for single atoms in~\cite{PhysRevA.97.053803}

We simulate these effects using \texttt{Pulser} and \texttt{EMU-MPS}~\cite{Silv_rio_2022,bidzhiev2025efficientemulationneutralatom}, and additionally account for the impact of thermal motion on the many-body Rydberg interactions following the approach of Refs.~\cite{Scholl_2021,Shaw_2024}. Finite-temperature effects are included by sampling atomic configurations at $T = 20\,\mu\mathrm{K}$, thereby capturing the influence of thermal motion on both single-particle parameters and interaction strengths. Dephasing is modeled phenomenologically with a coherence time $T_{\mathrm{deph}} = 20\,\mu\mathrm{s}$ (corresponding to a rate $\gamma_\phi = 0.05\,\mu\mathrm{s}^{-1}$). Fluctuations in the control parameters are included as static shot-to-shot disorder: the Rabi frequency $\Omega$ is sampled with a relative standard deviation of $1\%$, while the detuning $\delta$ is sampled with an absolute standard deviation $\sigma_\delta = 0.31\,\mathrm{rad}/\mu\mathrm{s}$. Spontaneous decay from the Rydberg state is neglected, as the characteristic lifetime is much longer than the experimental evolution time of $4\,\mu\mathrm{s}$. SPAM errors are characterized by detection error probabilities $\varepsilon = 1\%$ and $\varepsilon' = 7\%$, together with a preparation error $\eta \approx 1\%$; unless otherwise stated, these are not explicitly included in the coherent time-evolution model.

This noise model is implemented by sampling error parameters from experimentally motivated and testable distributions, followed by the execution of multiple Monte Carlo trajectories~\cite{noisemodel}. These factors together provide a coherent explanation for the observed deviations and highlight the utility of noise-aware modeling in interpreting quantum hardware behavior, see Fig~\ref{Sup_Fig2}(a-b). 

In simulations of the TFIL, the spectrum is highly sensitive to calibration uncertainties, such that even small calibration errors can shift the entire Fourier spectrum and spoil quantitative agreement with classical numerics; to account for this, we include a $1\%$ reduction in the atomic spacing and a $5\%$ increase in the detuning strength, which is within the uncertainty range of residual systematic offsets of the device, and which leads to much better quantitative agreement with the QPU data.; although the relevant symmetry is not strictly present away from the critical point, we find that its associated frequency ratios nonetheless accurately predict the peaks observed in the shifted spectrum. By carefully characterizing the experimental calibration and identifying the corresponding shift in the relevant energy scale, an increase of approximately $6\%$, we can therefore rescale the time axis of the analogue QPU data, bringing it into agreement with the exact theoretical results, See Fig.~\ref{Sup_Fig2}(c). Overall, we find that the experimentally calibrated noise levels are sufficiently low to allow access to the dynamical regimes of interest studied in this work.
\end{widetext}
\end{document}